\documentclass{aa}
\usepackage{graphicx,epsfig}

\def\mass#1{${\mathrm{#1\, M}_\odot}$}
\newcommand{\chem}[2]{$\rm{}^{#1}\kern-0.8pt#2$}
\newcommand{\chim}[2]{\rm{}^{#1}\kern-0.8pt#2}
\newcommand{\reac}[6]{$\rm\,{}^{#1}\kern-0.8pt{#2}\,({#3}\,,{#4})\,
           {}^{#5}\kern-0.8pt{#6}\,$}

\begin{document}


\title{Nucleosynthesis of s-elements in zero-metal AGB stars}

\author{S. Goriely and L. Siess} 
\institute{Institut d'Astronomie et d'Astrophysique, Universit\'e Libre de Bruxelles,
CP 226, B-1050 Brussels, Belgium}

\date{Received date; accepted date}

\abstract{Contrary to previous expectations, recent evolutionary models of
zero-metallicity stars show that the development of mixing episodes
at the beginning of the AGB phase allows low- and intermediate-mass stars
to experience thermal pulses. If these stars, like their metal-rich
counterparts, also experience partial mixing of protons from the H-rich
envelope into the C-rich layers at the time of the third dredge-up, an
extensive neutron capture nucleosynthesis leads to the
production of s-process nuclei up to Pb and Bi. Nucleosynthesis
calculations based on stellar AGB models are performed assuming a
parameterized H-abundance profile below the convective envelope at the time
of the third dredge-up. Despite the absence of Fe-group elements, the large
neutron flux resulting from the \reac{13}{C}{\alpha}{n}{16}{O} reaction
leads to an efficient production of s-process elements starting from the
neutron captures on the C-Ne isotopes.  Provided partial
mixing of protons takes place, it is shown that
population III AGB stars should be enriched in s-process elements and overall in Pb and
Bi.  
}

\titlerunning{s-process in  zero-metal AGB stars}
\maketitle

\section{Introduction}

The determination of chemical abundances at the surface of evolved stars
provides important clues on their structural evolution and traces the
hidden processes taking place in their nuclearly active interior.
Asymptotic giant branch (AGB) stars, and in particular zero-metallicity
stars, represent an ideal laboratory for such a study. They correspond to
the late phase of the evolution of stars with masses between about 1 and
\mass{8}, which includes more than 80\% of all stars; they exhibit peculiar
surface abundances as compared to other red giant stars; and many of them
are characterized by a strong mass loss (up to \mass{10^{-4}}~yr$^{-1}$)
which eject the surface material into the interstellar medium, contributing
thereby to the galactic chemical evolution. Among those, the population III
stars, i.e the first to form with the chemical composition inherited from
the Big Bang nucleosynthesis, are of particular interest to understand the
early enrichment of the universe in elements heavier than hydrogen and
helium. Little is known observationally on population III stars, since so
far no star of population III initial composition has yet been found.  If
s-process enriched AGB stars are commonly observed nowadays, they all
correspond to metallicities [Fe/H]~$\ga -3$. However, the detection of
carbon and nitrogen enrichments in the intergalactic medium at high
redshifts and in a considerable number of extremely metal-poor stars of our
Galaxy suggests the existence of primordial AGB stars (see e.g Abia et
al. 2001).

The abundance peculiarities observed at the surface of AGB stars are
understood as resulting from the mixing of material synthesized in the
interior with the surface layers as the convective envelope deepens during
the so-called third dredge-up (denoted hereafter 3DUP). But the
nucleosynthesis of s-process elements requires neutron-rich environments
that cannot be accounted for by the sole action of the dredge-up.  One way
to solve the problem is to assume the partial mixing (PM) of protons from
the envelope into the \chem{12}{C}-rich layers during a 3DUP event (Iben \&
Renzini 1982, Busso et al.  1999). The incomplete operation of the CN cycle
forms a ``\chem{13}{C} pocket'' which burns radiatively and releases a
large number of neutrons. However, AGB models are still subject to large
uncertainties concerning the consistent prediction of both the 3DUP and PM
processes (e.g Mowlavi 1999). In particular, the 3DUP and PM properties are
sensitive to the stellar characteristics (such as the mass, metallicity or
mass loss rate) and to the numerical scheme used to describe the convective
boundaries (e.g Frost \& Lattanzio 1996). In the past years, the mixing
scenario has been revived by the incorporation of new physical ingredients
to model overshooting (Herwig et al. 1999) and rotation (Langer et
al. 1999). Models including these new treatments predict the formation of a
\chem{13}{C} pocket in a zone comprising some \mass{10^{-5}}.  However, the
total amount of \chem{13}{C} produced in this way depends on the PM
parameters (extent and efficiency of the overshoot, amplitude of rotation)
which cannot be quantified {\sl ab initio} at the present time. In this
context, the resulting uncertainties on the subsequent nucleosynthesis can
only be handled at the moment in a parametric way.

Recently several works have been devoted to the study of the structure and
evolution of primordial stars (Siess et al. 2001, Chieffi et al. 2001,
Marigo et al. 2001).  However, from a nucleosynthesis point of view, the
production of elements heavier than Mg and ultimately of iron in low- and
intermediate-mass population III stars remains largely a mystery. Assuming
that primordial AGB stars suffer PM of protons as their higher metallicity
counterparts do, we analyze its effect on the resulting nucleosynthesis
applying the PM model developed by Goriely and Mowlavi (2000) to the
realistic 1-D AGB models computed by Siess et al. (2001). The corresponding
modeling is described in Sect.  2. In Sect. 3, we study in detail the
nucleosynthesis operating in the PM zone and its consequences on the
surface abundance predictions of zero-metallicity AGB stars. In particular,
it is shown that the initial absence of Fe does not prevent the formation
of the highest-mass elements and leads to the overproduction of Pb and Bi.

\section{Astrophysical model and input physics}

The stellar models used in this study are part of a recent grid computed by
Siess et al. (2001) and we refer the reader to this publication for ampler
details. Among the available mass tracks, we selected the \mass{3} model
for two main reasons: it is our most advanced model which accounts for 22
thermal pulses and \mass{3} is a somewhat ``representative'' mass for an
AGB star, between the low- (\mass{1}) and massive (\mass{8}) ones. Briefly,
the evolution of this model (which lasts for about 250~Myr) proceeds as
follows: during the main sequence, the star cannot activate the CNO cycle
because of the absence of \chem{12}{C}. Consequently, it contracts and
develops higher central temperatures to sustain its energy budget by the
poorly efficient p-p chains. When the central temperature reaches $10^8$K,
the 3$\alpha$ reactions start and carbon is produced in the core.  The CNO
cycle finally activates when the \chem{12}{C} mass fraction exceeds $\sim
10^{-11}$. At the exhaustion of hydrogen in the center, the 3$\alpha$
reactions are still effective and the star enters the core He burning phase
without experiencing the first dredge-up. The beginning of the AGB phase of
zero-metallicity stars is characterized by the occurrence of mixing
episodes which enables the carbon produced in the He burning shell to mix
with the envelope. This carbon pollution of the structure activates the CNO
cycle in the H burning shell and allows primordial stars to resume a
``standard'' AGB evolution comparable to that of more metal rich
stars. Note that without this mixing episode, the \mass{3} models would
probably not suffer thermal pulses (and 3DUP). Primordial AGB stars are
characterized by massive CO cores as a result of initially expanding their
nuclear core to compensate for the non-activation of the CNO cycle. They
also rapidly undergo third dredge-up events (after 9 pulses in this case)
and exhibit higher temperatures in the pulse compared to more metal-rich
stars. Of nucleosynthetic relevance is the activation of very efficient hot
bottom burning (HBB) after the 20th pulse, where the temperature at the
base of the convective envelope reaches $\sim 8\times 10^7$K. The impact of
HBB on the surface abundances of CNO elements is discussed in details in
Siess et al. (2001). Finally, because of the low metal content in the
envelope, mass loss is rather inefficient and by the end of the
computations only $10^{-3} M_\odot$ is lost. Nevertheless, the successive
dredge up episodes will increase little by little the opacity in the
envelope and consequently the mass loss efficiency.

The temperature and density profiles given by the stellar evolution code
are used as input data to compute the detailed nucleosynthesis resulting
from the PM of protons.  Protons from the envelope are ingested
artificially from the bottom of the convective envelope into the underlying
C-rich region at the time of the 3DUP events.  We adopt two different
H-abundance profiles. The first one decreases exponentially with depth and
ranges from $X_{\rm p}^{\rm mix} = 0.7$ at the bottom of the convective
envelope to $10^{-6}$ at the bottom of the PM zone. The second H abundance
profile decreases slower with depth than the previous case. It corresponds
to the ``Slow'' curve in Fig.~10 of Goriely and Mowlavi (2000). However,
since the nucleosynthesis calculations are found to be almost identical in
both cases, we present here only results using the second H abundance
profile which can be considered as more characteristic of diffusion
processes (Herwig et al. 1999). We consider a PM zone of extension
$\lambda_{\rm pm}=0.05$ corresponding to 5\% in mass of the pulse-driven
convection zone.  This corresponds to a zone comprising some \mass{10^{-5}}
subject to the s-process neutron irradiation.

The nucleosynthesis resulting from the ingestion of the protons into the
C-rich layers at the time of each 3DUP is followed during the first 22
interpulse and pulse phases of the \mass{3} stellar model. In contrast to
Goriely and Mowlavi (2000), the ashes of the interpulse and pulse
nucleosynthesis are reingested in the next interpulse/pulse sequence, so
that multiple irradiations on former nucleosynthetic products are
consistently taken into account. The nuclear reaction network includes 817
nuclei up to Po, and all relevant nuclear, weak and electromagnetic
interactions. Note that the network needs to be extended to the
neutron-rich region because of the very large neutron densities generated
in the late He-flashes. The latest experimental and theoretical reaction
rates are taken from the Nuclear Astrophysics Library of the Brussels
University (available at http://www-astro.ulb.ac.be). More details can also
be found in Goriely and Mowlavi (2000).

\section{The neutron capture nucleosynthesis}

\begin{figure}[h]
\centerline{\epsfig{figure=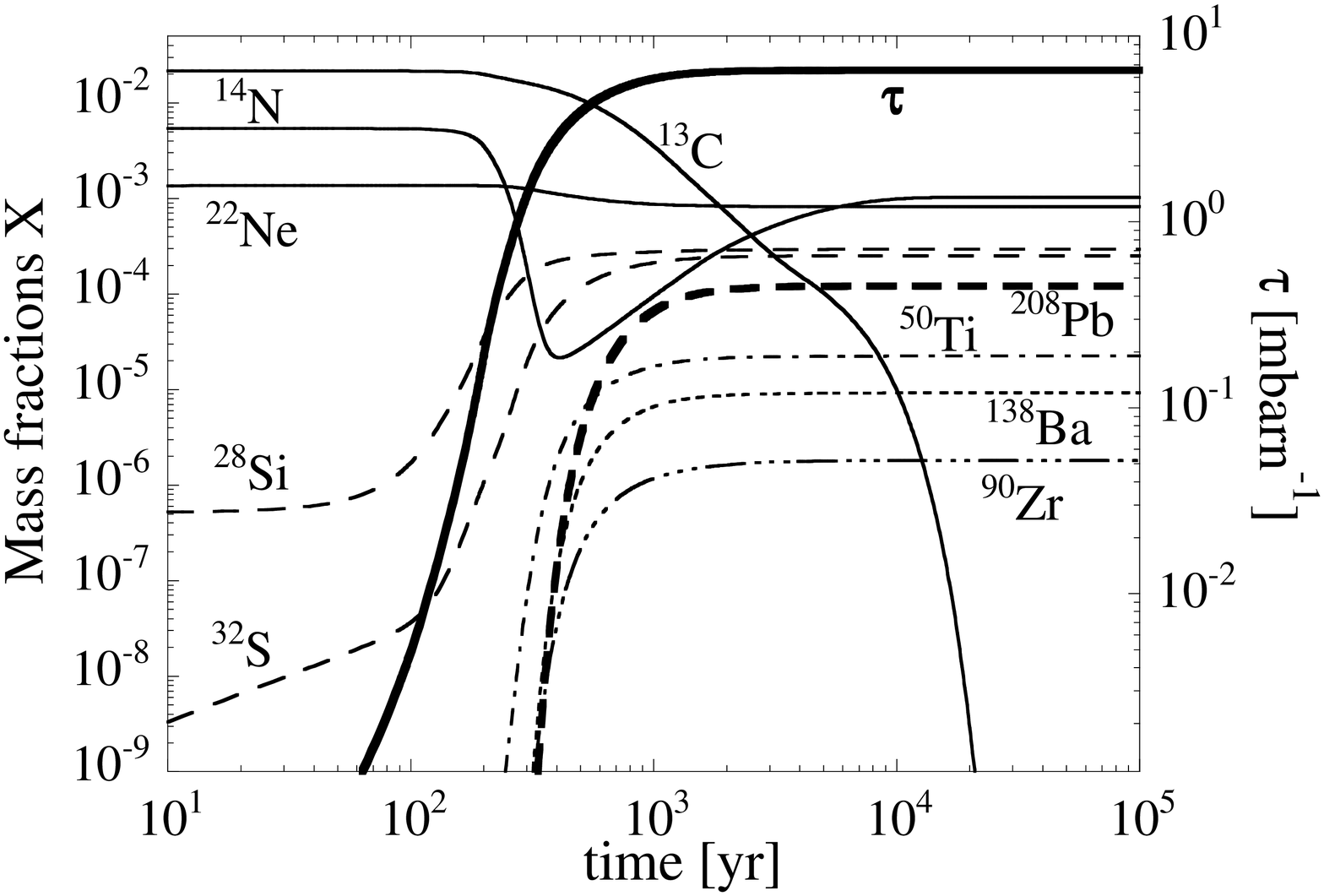,height=5.5cm,width=8cm}}
\caption{Mass fraction of some elements representative of the neutron
capture nucleosynthesis taking place in the PM layer characterized by an
initial proton-to-\chem{12}{C} ratio of $10^{-2}$ during the 21st
interpulse phase of the $M=3~M_{\odot}, Z= 0$ AGB star. Also given (bold
solid line) is the time dependence of the neutron exposure $\tau$.}
\label{fig01}
\end{figure}

To analyse the nucleosynthetic impact of the PM of protons into the C-rich
region, we first consider one interpulse/pulse/3DUP phase. The richest
nucleosynthesis occurs during the interpulse phase. More specifically, for
low proton-to-carbon ratios [$Y_{\rm p}/Y(^{12}{\rm C})\la 1$, where $Y(A)
= X(A)/A$ is the abundance of $A$ by number], neutrons are efficiently
produced after \chem{13}{C} has been synthesized, and subsequently
destroyed by \reac{13}{C}{\alpha}{n}{16}{O} (Fig.~\ref{fig01}). In the
absence of Fe-group nuclei, neutrons are captured by all the abundant C, N,
O, F and Ne isotopes and enable the production of high mass elements. The
fast $(n,\alpha)$ reaction on \chem{33}{S} forms a major bottle-neck
braking down the s-process flow partially. As soon as the bottle-neck is
passed, radiative neutron captures followed by $\beta$-decays on unstable
nuclei lead to the synthesis of all medium and heavy elements up to Pb and
Bi. The \reac{33}{S}{n}{\alpha}{30}{Si} recycling reduces the total amount
of s-elements produced by a factor of about 5. The evolution of the mass
fraction of some representative nuclei (with a magic number of neutrons
$N=28, 50, 82$ and 126) is displayed in Fig.~\ref{fig01}. In the series of
neutron captures and $\beta$-decays characteristic of the s-process, iron
group nuclei do not play any particular role, in contrast to the classical
picture of the s-process where the neutron captures take place on the
abundant \chem{56}{Fe} mainly.

\begin{figure}
\centerline{\epsfig{figure=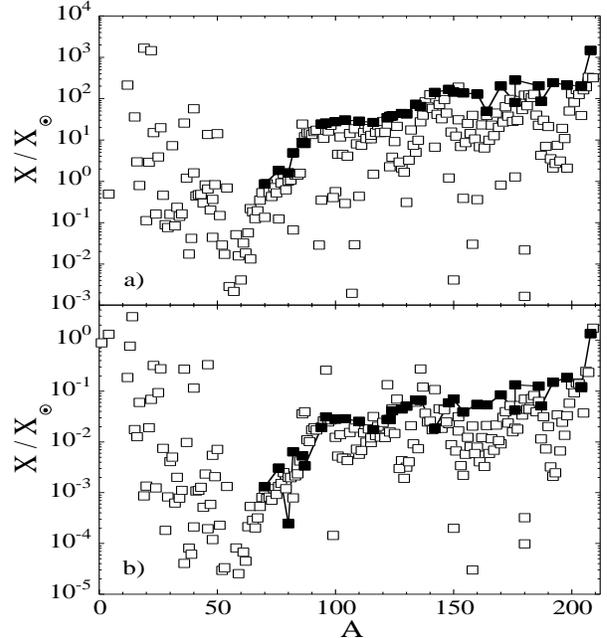,height=8.6cm,width=8cm}}
\caption{a) Abundance distribution (relative to solar) at the end of the 21st
interpulse phase (a) and following the next pulse (b) for the
$M=3~M_{\odot}, Z = 0$ AGB star. In panel (a), the distribution is averaged
over the mass range covered by the PM region (in this case $M_{\rm pm} = 4
\times 10^{-4}~M_{\odot}$) while case (b) gives the surface
abundance after the 3DUP. Full squares correspond to
s-only isotopes. Note the nuclear rearrangement of some s-only elements
during the pulse.}
\label{fig02}
\end{figure}

For $10^{-2} \la Y_{\rm p}/Y(^{12}{\rm C}) \la 1$, the neutron exposure
$\tau=\int_0^t N_{\rm n}~v_T~dt$ reaches 10~mbarn$^{-1}$ ($v_T$ is the most
probable relative neutron-nucleus velocity at temperature $T$) and the
neutron density $N_{\rm n}$ a few $10^9~\rm{cm}^{-3}$. These high values
explain the large production of metals up to Pb and Bi despite the absence
of initial iron (Fig.~\ref{fig02}a). For relatively high proton-to-$^{12}$C
abundance ratios [$Y_{\rm p}/Y(^{12}{\rm C})\ga 1$], \chem{14}{N} is more
abundant than \chem{13}{C} and prevents any significant neutron
production. In this case, the PM of protons only affects light species. The
fast \reac{14}{N}{n}{p}{14}{C} reaction is responsible for the production
of \chem{14}{C} and later of \chem{18}{O} by
\reac{14}{C}{\alpha}{\gamma}{18}{O}. This reaction also releases a large
amount of protons which are captured by \chem{18}{O} to form
\chem{15}{N}. Subsequently, in the pulse-driven convection zone,
\chem{19}{F} is largely produced by
\reac{15}{N}{\alpha}{\gamma}{19}{F}. However, as soon as envelope HBB is
activated, the dredged-up \chem{19}{F} is destroyed. HBB also increases the
surface abundance of \chem{14}{N} and \chem{23}{Na} significantly at the
expense of CNO isotopes and \chem{22}{Ne}, respectively.

\begin{figure}
\centerline{\epsfig{figure=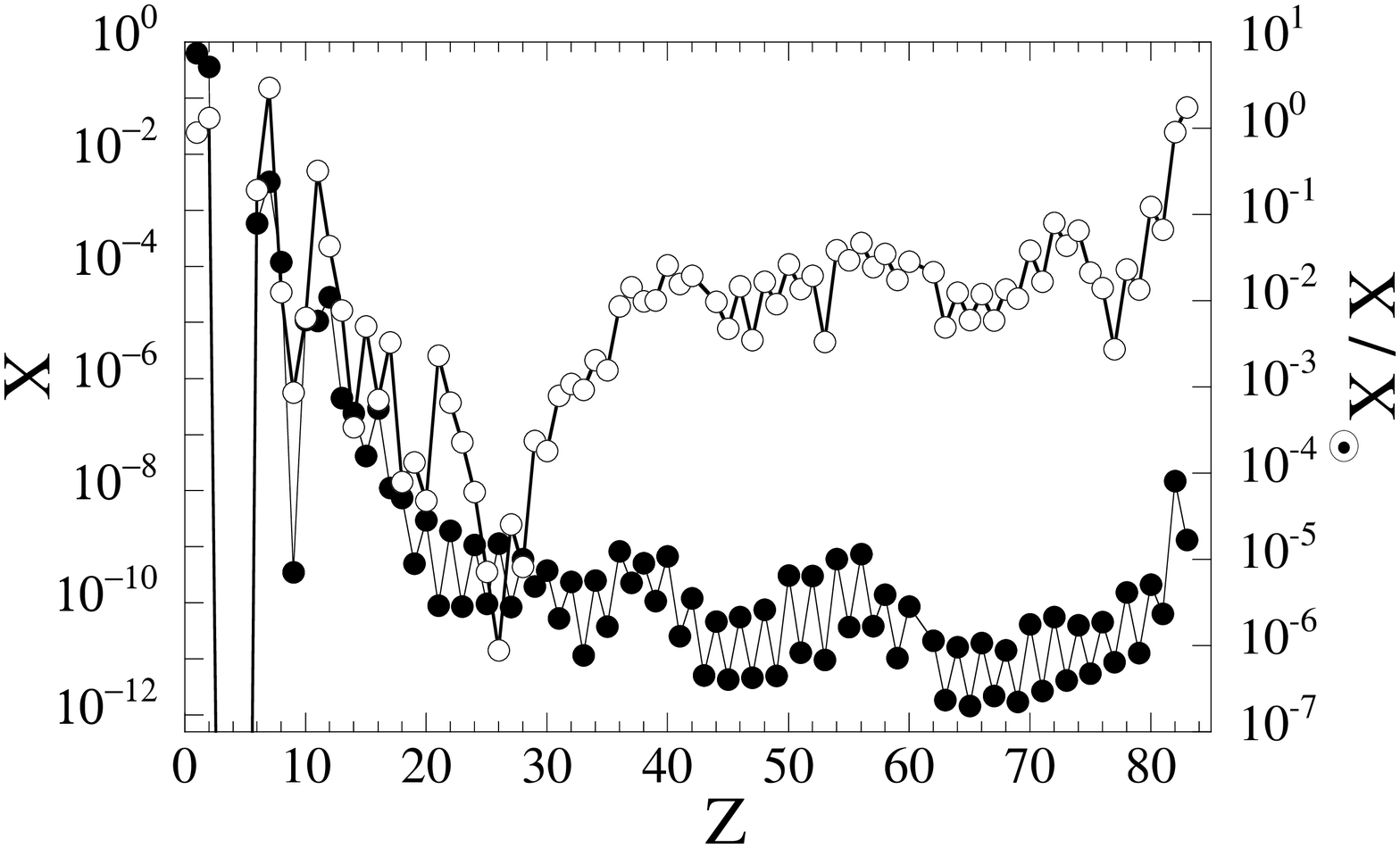,height=5.5cm,width=7.5cm}}
\caption{Elemental abundances (full circles) and
overabundances relative to solar (open circles) at the surface of the
$M=3~M_{\odot}$ zero-metallicity AGB star after the 22nd
convective pulse phase. }
\label{fig03}
\end{figure}

As the core mass grows, the temperature in the thermal pulse increases and
by the 20th pulse, it exceeds $4 \times 10^8$K. In these conditions,
\chem{22}{Ne} is rapidly burnt by \reac{22}{Ne}{\alpha}{n}{25}{Mg} and
additional neutrons are released. In the 22nd pulse, the neutron density
reaches $N_{\rm n}=10^{15}~{\rm cm}^{-3}$, but for a short period of time
and mostly near the bottom of the He-flash zone. Consequently, the
pulse-averaged resulting neutron exposure is small. Nevertheless, the
neutron burst allows for some rearrangements of the abundance distribution
of peculiar s-only isotopes (e.g \chem{80}{Kr} or $^{142}{\rm Nd}$, see
Fig.~\ref{fig02}b) and of some sr-nuclei by short-cutting the s-process
flow. In particular, \chem{122}{Sn}, \chem{136}{Xe}, \chem{137}{Ba} and
\chem{142}{Ce} are overproduced compared with neighboring
isotopes. Abundances of the so-called $T$-dependent s-branching nuclei, e.g
$^{164}{\rm Er}$, are also modified because of the high temperatures
prevailing in the pulse. Among the medium-mass nuclei, \chem{36}{S},
\chem{40}{Ar} and \chem{46}{Ca} are largely produced. Note that the large
dip in the $X/X_{\odot}$ scale observed in Figs.~\ref{fig02}-\ref{fig03}
for the Fe-group elements is due to their high solar values, and
consequently small abundances relative to solar.

Except for the first 6 interpulse/pulse sequences where the C abundance
left over by the preceding pulse-driven convective zone is small, the
nucleosynthesis resulting from the PM of protons is found to be
qualitatively identical from one interpulse/pulse phase to the other (the
only significant difference originates from the temperatures reached at the
bottom of the He-flash zone). The relatively small extent of the PM zone
($\lambda_{\rm pm}=0.05$) does not allow a significant cumulative neutron
irradiation on the ashes of the previously synthesized material. As a
consequence, the most abundant material, namely the elements comprised
between C and Ne, always remain the seed nuclei for the neutron captures in
the PM zone.

The above-described nucleosynthesis is in many respects similar to the one
found in the low-metallicity ($Z=0.001$) star subject to the same mixing of
protons (Goriely \& Mowlavi, 2000). In the radiative C-rich layers during
the interpulse phase, the overall abundance distributions are much the same
even if the seed nuclei at the origin of the heavy elements are in both
cases quite different. For stars with metallicities $Z \la 0.001$, the
number of neutrons available per seed nuclei exceeds the number needed to
produce the heaviest s-nuclei, i.e Pb and Bi, so that the type of seeds as
well as their initial abundances only affect the absolute abundance of
heavy elements synthesized by the s-process, but not their relative
distribution. However, for extremely low metallicity stars, the higher
temperatures reached at the bottom of the pulse-driven convective zones can
reshape the s-abundance distribution differently.

Provided partial mixing of protons occurs in population III stars in a
similar way as in more metal-rich stars, the primordial stars should
exhibit a strong overabundance of Pb and Bi compared with other s-process
elements ([Pb/s] $\ga 1$). Therefore, such stars could be dubbed population
III Pb-stars (Goriely \& Mowlavi 2000, Van Eck et al. 2001).  In addition
to their large overabundance in \chem{14}{N} (mainly due to the HBB
transmuting all the CNO envelope isotopes into \chem{14}{N}), these
Pb-stars are also predicted to have a large \chem{23}{Na} content, as shown
in Fig.~\ref{fig03}. Van Eck et al. (2001) observed low-metallicity stars
(down to [Fe/H]=-2.45) enriched in s-process elements and characterized by
large Pb overabundances ([Pb/hs]$\ga$ 1, where hs denotes heavy s-elements
such as Ba, La or Ce). The discovery of these Pb-stars confirms the
s-process efficiency in low-metallicity stars.  However, this conclusion is
not reached by Aoki et al. (2000) who found in a metal-poor s-process
enriched stars ([Fe/H]=-2.7), a Pb overabundance ([Pb/Fe]=+2.6) but no
Pb-to-s overabundance ([Pb/hs]$\simeq$0). This latter observation remains
to be explained.
 
 While the discovery of Pb-stars (Van Eck et al. 2001) strongly confirms
the ``proton-mixing'' scenario for the detailed operation of the s-process
in AGB stars, their is still no evidence neither on the existence of
population III stars nor on the possible mixing of protons in such
stars. For this reason, it would be premature at this stage to estimate the
possible contribution of population III stars to the galactic enrichment in
s-process elements, and more particularly in Pb (e.g Travaglio et
al. 1999).

\section{Conclusion} 
Assuming that population III stars experience, just like more metal-rich
stars, a partial mixing of protons from the H-rich envelope into the C-rich
layers at the time of the third dredge-up, an extensive neutron capture
nucleosynthesis is found to take place leading to the efficient production
of s-process nuclei. Despite the absence of Fe-group elements, the large
neutron flux resulting from the \reac{13}{C}{\alpha}{n}{16}{O} reaction
transmutes part of the C-Ne isotopes into s-nuclei with a large
overabundance of Pb and Bi. Provided primordial stars are observable today,
if they exhibit a surface enrichment in s-process elements, a strong
overabundance of Pb and Bi compared with other s-process elements ([Pb/s]
$\ga 1$) is predicted. These population III Pb-stars remain to be observed.
Their discovery or the indirect identification of their pollution in later
stellar generations would prove that the galactic enrichment in s-process
elements could have started already a few hundred million years after the
Big Bang.

\begin{acknowledgements}
S.G. is FNRS Research Associate. L.S. benefits of a European TMR ``Marie
Curie'' fellowship at ULB.
\end{acknowledgements}


\begin{thebibliography}{}

\bibitem[]{} Abia C., Dominguez I., Straniero O. et al., 2001, ApJ 557, 126



\bibitem[ ]{  }  Aoki W., Norris J.E. Ryan S.G., et al., 2000,
ApJ 536, L97


\bibitem[ ]{ } Busso M., Gallino R., Wasserburg G.J., 1999, ARA\&A 37, 239


\bibitem[2001]{chi01} Chieffi A., Dominguez I., Limongi M., Straniero
O. 2001, ApJ 554, 1159 


\bibitem[1996]{frost96} Frost C.A., Lattanzio J.C. 1996, ApJ 473, 383


\bibitem[ ]{  } Goriely S., Mowlavi N., 2000, A\&A 362, 599



\bibitem[ ]{  }  Herwig F., Bl\"ocker T., Sch\"onberner D., 1999, in 191st
IAU Symposium (PASP, eds. T. Le Bertre et al.), p.41

\bibitem[ ]{  } Iben I.Jr., Renzini A., 1982, A\&A 263, L23

\bibitem[ ]{  } Langer N., Heger A., Wellstein S., Herwig F., 1999, A\&A
346, L37 


\bibitem[2001]{Mar01} Marigo P., Girardi L., Chiosi C., Wood P.R. 2001,
A\&A 371, 152 

\bibitem[ ]{  } Mowlavi N., 1999, A\&A 344, 617.

\bibitem[ ]{  } Siess L., Livio M., Lattanzio J., 2001, ApJ, submitted


\bibitem[ ]{  } Travaglio C., Galli D., Gallino R. et al., 1999, ApJ 521, 691

\bibitem[ ]{ } Van Eck S., Goriely S., Jorissen A., Plez B., 2001, Nature
412, 793 

\end{thebibliography}
\end{document}